\documentclass[prd,twocolumn,showpacs,floatfix,superscriptaddress]{revtex4}

\usepackage{graphicx}
\usepackage{epsfig}
\usepackage{bm}
\usepackage[utf8]{inputenc}
\usepackage{amssymb}
\usepackage{float}
\usepackage{amsmath}
\usepackage{dcolumn}
\usepackage{cancel}
\usepackage[colorlinks]{hyperref}
\usepackage[usenames,dvipsnames]{color}
\hypersetup{
     breaklinks=true,
    pdfstartview={FitH},    
    colorlinks=true,       
    linkcolor=blue,          
    citecolor=red,        
    filecolor=magenta,      
    urlcolor=blue,           
    anchorcolor=green,      
    linktocpage=true
}

\def\doi{http://doi.org}
\def\arxiv{http://arxiv.org/abs}

\begin{document}

\title{Power-law holographic dark energy and cosmology}

\author{Eirini C. Telali}
 
\affiliation{National Observatory of Athens, Lofos Nymfon, 11852 Athens, 
Greece}

\author{Emmanuel N. Saridakis}
\affiliation{National Observatory of Athens, Lofos Nymfon, 11852 Athens, 
Greece}
\affiliation{CAS Key Laboratory for Researches in Galaxies and Cosmology, 
Department of Astronomy, University of Science and Technology of China, Hefei, 
Anhui 230026, P.R. China} 
\affiliation{Department of Astronomy, School of Physical Sciences, 
University of Science and Technology of China, Hefei 230026, P.R. China}

\begin{abstract}  
We formulate  power-law  holographic dark energy, which is a modified 
holographic dark energy model based on the extended entropy relation arising  
from the consideration of  state mixing between the ground and the excited ones  
in the calculation of the entanglement entropy. We construct two cases of the 
scenario, imposing the usual future event horizon  choice, as well 
as the Hubble one. Thus, the former model is  a one-parameter extension of 
standard holographic dark energy, recovering it in the limit where power-law 
extended entropy recovers Bekenstein-Hawking one, while the latter   belongs to 
the class of running vacuum models, a feature that  may reveal the 
connection between holography and the   renormalization 
group running. For both models  we 
extract the differential equation that determines the evolution of the 
dark-energy density parameter and we provide the expression for the 
corresponding equation-of-state parameter. We find that the scenario can 
describe the sequence of epochs in the Universe evolution, namely the domination 
of matter followed by the domination of dark energy. Moreover, the dark-energy 
equation of state presents a rich behavior, lying 
in the quintessence regime or passing into the phantom one too, depending on 
the values of the two model parameters, a behavior that is richer than the one 
of standard 
holographic dark energy.

\end{abstract}


\maketitle

\section{Introduction}

According to accumulating observational evidence, the Universe has    
exhibited the transition to the accelerated era in the recent cosmological past.
One can attribute this unexpected  feature  in the presence of the cosmological 
constant, however the corresponding problem related to the smallness of its 
observed value, as well as the possibility that the acceleration has  a 
dynamical nature, has led to a variety of other possibilities. A first avenue 
of modification  of the standard cosmological paradigm is to introduce the 
concept of dark energy 
\cite{Copeland:2006wr,Cai:2009zp,Bamba:2012cp}. The second road that one can 
follow is to modify the gravitational theory itself, resulting to theories that 
exhibit richer structure and phenomenology that could eventually lead to an 
accelerating Universe  
\cite{Nojiri:2010wj,Capozziello:2011et,Cai:2015emx,CANTATA:2021ktz}.

Holographic dark energy is a different alternative that can provide the 
explanation for the Universe acceleration. The basis of this scenario is 
the holographic principle 
\cite{tHooft:1993dmi,Susskind:1994vu,Bousso:2002ju}, applied in a   
cosmological   framework
\cite{Fischler:1998st,Bak:1999hd,Horava:2000tb}.
In particular, since the largest distance of a quantum field theory, required 
for its applicability at large distances, is related to its ultra-violet (UV) 
cutoff, which is 
in turn  related 
to the vacuum energy  \cite{Cohen:1998zx,Addazi:2021xuf}, one can attribute a
holographic origin to the vacuum energy, which will then be a
  form of  holographic dark energy \cite{Li:2004rb,Wang:2016och}.
    Holographic dark energy has been shown to exhibit   very 
interesting cosmological phenomenology
\cite{Li:2004rb, 
Wang:2016och,Horvat:2004vn,Huang:2004ai,Pavon:2005yx,
Wang:2005jx,
Nojiri:2005pu,Kim:2005at, Setare:2006wh,Setare:2008hm, 
Sheykhi:2009dz, 
Li:2009bn,Zhang:2009un,Micheletti:2010cm,Lu:2009iv,
Micheletti:2009jy,Aviles:2011sfa,Luongo:2017yta}, and additionally a large 
amount of research has been 
devoted to its extensions.

There are two main ways of extending the basic scenario of holographic dark 
energy. The first is to apply  different horizons for the Universe, such as the 
event horizon, the apparent horizon, the age of the universe, the conformal 
time, the inverse square root of the Ricci curvature  and the Gauss Bonnet 
term, etc \cite{Gong:2004fq,Saridakis:2007cy, Cai:2007us,
Setare:2008bb,Gong:2009dc,Suwa:2009gm,
Jamil:2010vr,BouhmadiLopez:2011xi,Malekjani:2012bw,Khurshudyan:2014axa,
Landim:2015hqa,Shekh:2021ule}. On the other hand, the 
second possibility of modification is the consideration of a modified entropy 
expression for the Universe horizon instead of the standard   Bekenstein-Hawking 
one, such as Tsallis entropy, Barrow entropy, Kaniadakis entropy, logarithmic 
corrected entropy etc
\cite{Pasqua:2015bfz,Jawad:2016tne,Pourhassan:2017cba,
Saridakis:2017rdo,Nojiri:2017opc,Saridakis:2018unr, DAgostino:2019wko,
Saridakis:2020zol,Dabrowski:2020atl,
daSilva:2020bdc, Anagnostopoulos:2020ctz, 
Mamon:2020spa,Bhattacharjee:2020ixg,Huang:2021zgj,
Drepanou:2021jiv,Hossienkhani:2021emv,
Nojiri:2021iko,Jusufi:2021fek,Hernandez-Almada:2021aiw,Hernandez-Almada:2021rjs,
Luciano:2022pzg} .
 
Attributing an entropy to a horizon, and in particular relating it to its area, 
is connected    with the topological restrictions 
of quantum interactions of quantum states and emerges from seemingly distinct 
areas in physics, such as black holes, quantum field theory, information 
theory, and 
quantum many-body physics \cite{Eisert:2008ur}. 
The first to suggest an area-law 
dependence of the black hole entropy was Bekenstein \cite{Bekenstein:1973ur}, 
who, 
based on information theory and dimensional arguments, derived the well known 
 formula, that was later specified by Hawking 
\cite{Hawking:1975vcx}, namely the Bekenstein-Hawking entropy  $S_{BH}= 
\frac{A}{4 l_{P}^{2}}$,
where $A$ is the horizon area and $ l_{P}$ the Planck length, in units  
$k_{_B}=c=\hbar=1$. This kind of 
behaviour was verified as compatible with the 
notion of quantum gravity through the holographic principle.   An identical 
entropy behaviour reproducing the area 
law, was also derived by Bombelli \cite{Bombelli:1986rw} and later by Srednicki 
\cite{Srednicki:1993im}, for the entanglement entropy of a spherical region in 
the 
case of scalar massless fields, and this concurrence is not thought to be 
accidental. Indeed, from the recent developments in AdS/CFT correspondence 
\cite{Maldacena:1997re} and 
the Ryu-Takayanagi formula \cite{Ryu:2006bv}, to the concept of emergent 
gravity \cite{Sakharov:1967pk}, indications appear to agree on a deeper 
connection 
among the manifestations of the area law.

In the calculation  of the entanglement entropy one starts from the vacuum 
ground state. However,   adding the excited states one obtains  
  a power-law correction   in the entanglement entropy. In 
particular, considering a 1-particle excitation one results to a modified 
entropy relation of the form   \cite{Das:2007mj,Radicella:2010ss}
\begin{equation} \label{Power_corr_Entropy}
	S = S_{BH} + c\left(\frac{A}{\epsilon^2}\right)^{-\gamma},
\end{equation}
where the exponent $\gamma$ depends on the degree 
of mixing  of the ground state with the excited one, varying between -1 and 
0, and $c$ is a constant. Furthermore, $\epsilon$ is the UV cutoff of the 
scalar field theory used to derive the entropy, and in the following we will set 
it equal to    the Plank length $l_p = M_p^{-1}$, where $M_p$ the 
Plank mass. Hence, the above expression for $c=0$ recovers the standard 
Bekenstein-Hawking 
entropy, while for   $\gamma = 0 $  we obtain  the topological entropy 
constant term \cite{Kitaev:2005dm}. 
  
In this work we will apply the power-law entropy relation in order to formulate 
 holographic dark energy, and study its cosmological implications. Although in 
the literature there were some attempts towards this direction, they remained 
incomplete with complete absence of cosmological applications 
\cite{Sheykhi:2011egx,Khodam-Mohammadi:2011oji,Ebrahimi:2010xz,Pasqua:2012hm}.
The plan of the manuscript is the following: In Section   \ref{model1} we 
formulate power-law holographic dark energy,  presenting the corresponding 
cosmological equations and   extracting the expressions for the dark 
energy density and equation-of-state parameters, focusing on two cases of 
the horizon choice. In Section 
\ref{Cosmolevolut} we investigate the resulting 
cosmological behavior. Lastly, in Section \ref{Conclusions}   we summarize the 
obtained results.

\section{Power-law holographic dark energy}
\label{model1}

  In this section we  formulate power-law  holographic 
dark energy. We start be recalling that  holographic dark energy arises from the
 inequality $\rho_{DE} 
L^4\leq S$, with $L$   the Universe horizon and $S$ the black-hole entropy 
relation applied for $L$ 
\cite{Li:2004rb,Wang:2016och}. If we use the standard
Bekenstein-Hawking 
entropy 
$S_{BH}\propto A  M_p^2/4=\pi  M_p^2 L^2$,   
then the 
saturation of the previous inequality provides standard holographic dark 
energy, 
i.e.  $\rho_{DE}=3c_1^2 M_p^2 L^{-2}$, with $c_1$ the 
model parameter.
     
However, as we mentioned  in the Introduction, there are many modifications of 
Bekenstein-Hawking  entropy. One of them is the power-law extended entropy 
(\ref{Power_corr_Entropy}), which arises from the consideration of     the 
mixing of excited states with the ground one  in the entanglement 
entropy \cite{Das:2007mj}. Re-expressing it as 	$S = S_{BH} + c 
S_{BH}^{-\gamma}$ and inserting it in the inequality
 $\rho_{DE} 
L^4\leq S$  we acquire 
  \begin{equation}
\label{rhoDE}
\rho_{DE}=  c_1 \pi M_{p}^{2} L^{-2} + c_2 \left(\pi 
M_{p}^{2}\right)^{-\gamma} L^{-2\gamma-4},
\end{equation}
  	where $c_1$ and $c_2$ are dimensionless constants (the   
parameter $c$ has been 
absorbed in $c_2$)  and thus in units 
$k_{_B}=c=\hbar=1$ the energy density has units of $[M]^4$ as expected. 
Therefore, as mentioned above, for 
$c_2=0$ the above expression becomes the 
standard holographic dark energy  $\rho_{DE}=3c_1^2 M_p^2 L^{-2}$.  
 
 As usual, we  consider  a flat  
Friedmann-Robertson-Walker (FRW) metric of the form
\begin{equation}
\label{FRWmetric}
ds^{2}=-dt^{2}+a^{2}(t)\delta_{ij}dx^{i}dx^{j}\,,
\end{equation}
with $a(t)$ the scale factor. 
Given this geometry we have two choices for  the maximum  length   $L$ that 
appears in the above relations.
In 
the case of usual holographic dark energy   one can show
 that  $L$ cannot be the Hubble 
horizon $H^{-1}$ (with $H\equiv \dot{a}/a$   the Hubble function), since 
 the resulting model cannot lead to acceleration
  \cite{Hsu:2004ri}. Therefore, the standard choice is 
 the   future event horizon, defined as
  \cite{Li:2004rb} 
\begin{equation}
\label{futrhoriz}
R_h\equiv a\int_t^\infty \frac{dt}{a}= a\int_a^\infty \frac{da}{Ha^2}.
\end{equation}
Hence, in the following we will use $R_h$ as the IR cutoff of the theory.
However, since the present scenario is a novel one,  we do have the choice to 
use  the Hubble 
horizon $H^{-1}$ as $L$, since the resulting model can indeed describe 
acceleration.  In the following subsections we will study these two cases 
separately.

\subsection{Model 1: Holography on the future event horizon}

Let us use the  future event horizon $R_h$ of (\ref{futrhoriz}) as the 
infra-red (IR) 
cutoff. In this case, the holographic dark energy density (\ref{rhoDE}) will be 
    \begin{equation}
\label{rhoDEs}
\rho_{DE}=  c_1 \pi M_{p}^{2} R_h^{-2} + c_2 \left(\pi 
M_{p}^{2}\right)^{-\gamma} R_h^{-2\gamma-4}.
\end{equation}
Moreover, the two Friedmann equations in a universe containing the 
aforementioned holographic dark energy, as well as the matter 
perfect fluid, are written as
 \begin{eqnarray}
\label{Fr1b}
3M_p^2 H^2& =& \ \rho_m + \rho_{DE}    \\
\label{Fr2b}
-2 M_p^2\dot{H}& =& \rho_m +p_m+\rho_{DE}+p_{DE},
\end{eqnarray}
with $p_{DE}$ the pressure of  the power-law holographic dark energy, and 
$\rho_m$, $p_m$  the matter
   energy density and pressure     respectively. Note that we can additionally 
consider the  matter
conservation equation  
\begin{equation}\label{rhoconserv}
\dot{\rho}_m+3H(\rho_m+p_m)=0,
\end{equation}
which according to Friedmann equations then implies the dark-energy conservation
equation 
\begin{equation}
\label{rhodeconserv}
\dot{\rho}_{DE}+3H(\rho_{DE}+p_{DE})=0.
\end{equation}
 
 We proceed by introducing     the density parameters of the  dark energy and 
matter sector as
 \begin{eqnarray}
 && \Omega_m\equiv\frac{1}{3M_p^2H^2}\rho_m
 \label{Omm}\\
 &&\Omega_{DE}\equiv\frac{1}{3M_p^2H^2}\rho_{DE}.
  \label{ODE}
 \end{eqnarray}
Furthermore, inserting (\ref{ODE}) into (\ref{rhoDEs}) gives
	\begin{equation}\label{ODE_equation}
		\Omega_{DE}H^{2} = C_1 R_h^{-2} + C_2 R_h^{-2\gamma-4},			
	\end{equation}	
	with  $C_1 \equiv \frac{c_1 \pi}{3}$ and $C_2 
\equiv\frac{c_{2} \pi^{-\gamma} 
M_{p}^{-2+2\gamma}}{3} $   the model parameters.
On the other hand, differentiating  (\ref{futrhoriz}) with respect to time
we obtain the useful relation $\dot{R}_h=HR_h-1$, and if we use $x=\ln a$ as 
the independent variable it gives rise to
	\begin{equation}\label{Rhderiv}
 R_h'=R_h-H^{-1},
	\end{equation}	
 with primes 
denoting derivatives with respect to $x$.
Hence, taking the $x$-derivative of  (\ref{ODE_equation}) and using 
(\ref{Rhderiv}), we finally obtain
	 \begin{eqnarray} 
	 		&&  
	 		\tilde{\Omega}(F'+ \tilde{F}' + 2^{\frac{1}{3}}Q') -		
	\tilde{\Omega}'(F+ \tilde{F} + 2^{\frac{1}{3}}Q) \nonumber\\
	&&- \tilde{\Omega}(F+ 
\tilde{F} + 
2^{\frac{1}{3}}Q) +  3 H^{-1} 2^{\frac{1}{3}}  \tilde{\Omega}^2  =0,
\label{difequation}
  	 \end{eqnarray} 
	where 
	\begin{eqnarray}
&					&Q = (2\gamma+4)\Omega_{DE}H ,  \nonumber\\
			&	&\tilde{\Omega} = (\Omega_{DE} H^2)' + H Q ,\nonumber\\
						&&	A =  18C_{1}(\gamma+1)\tilde{\Omega} (1+3 
H^{-1}\tilde{\Omega})
+2Q^3 , \nonumber\\
			&	&B = -6C_{1}(\gamma+1)\tilde{\Omega}-Q^2,  \nonumber\\
			&	&F = (\sqrt{A^2+4B^3}+A)^{\frac{1}{3}}, \nonumber\\
			&&	\tilde{F} = (-\sqrt{A^2+4B^3}+A)^{\frac{1}{3}}.
			\label{definitions}
 	\end{eqnarray}
Note that the derivatives of $H$ in the above formulas can be expressed in 
terms of derivatives of $\Omega_{DE}$ through the simple relation
 	\begin{equation}\label{Hrel2}
		\frac{1}{Ha}=\frac{\sqrt{a(1-\Omega_{DE})}}{H_0\sqrt{\Omega_{m0}}},
	\end{equation}
which is just the first Friedmann 
equation (\ref{Fr1b}) in the case of   dust matter, i.e $p_m=0$, in which 
case (\ref{rhoconserv}) leads to $\Omega_m=\Omega_{m0}/a^3$, with the 
subscript ``0'' denoting the present values of a quantity.

In summary, equation  (\ref{difequation}) is a second-order differential 
equation for  
	 $\Omega_{DE}$ as 
a function of $x=\ln a$, and hence it 
determines the 
evolution of power-law holographic dark energy. 
As expected, in the case $C_2=0$ it   recovers   
the corresponding differential equation of standard holographic dark energy  
\cite{Li:2004rb},  
namely
$\Omega_{DE}' = 
\Omega_{DE}(1-\Omega_{DE})\left(1+2\sqrt{\frac{3M_p^2\Omega_{DE}}{3 
c_1^2 M_p^2}}
\right)$, 
 which is solved in an  implicit form  \cite{Li:2004rb}. 
 
Concerning the important observable $w_{DE}\equiv p_{DE}/\rho_{DE}$, 
namely the dark-energy equation-of-state parameter, from the conservation 
equation (\ref{rhodeconserv}) we obtain $\rho_{DE}'+3\rho_{DE}(1+w_{DE})=0$, 
which using the density parameter and the definitions 
    (\ref{definitions}) gives
    \begin{eqnarray}
\label{wdedefin}
 w_{DE}=-1+\frac{HQ-\tilde{\Omega}}{3H^2\Omega_{DE}}.
\end{eqnarray}
 Thus,    knowing 
$\Omega_{DE}$ from  (\ref{difequation})  
allows us to calculate   
    $w_{DE}$ as a function of $\ln a$.
    Once again we mention that for $C_2=0$ 
 expression  (\ref{wdedefin})
 provides  the result of standard 
holographic 
dark energy, that is $w_{DE} 
=-\frac{1}{3}-\frac{2}{3}\frac{\sqrt{\Omega_{DE}}}{c}$ 
\cite{Wang:2016och}, as expected. 
Note that  $w_{DE}$ can be   quintessence-like or 
phantom-like according to the model parameter choices.
Lastly,  it proves convenient to introduce the deceleration 
parameter  
  \begin{equation}
  \label{qdeccel}
q\equiv-1-\frac{\dot{H}}{H^2}=\frac{1}{2}+\frac{3}{2}\left(w_m\Omega_m+w_{DE}
\Omega_{DE}
  \right),
\end{equation}
which  for  dust matter ($w_m=0$)  can be straightforwardly found if
$\Omega_{DE}$ (and therefore $w_{DE}$) is known.

\subsection{Model 2: Holography on the Hubble horizon}

We proceed to the construction of a different scenario, which arises from the 
use of the Hubble horizon $H^{-1}$ as the IR 
cutoff. In this case, the holographic dark energy density (\ref{rhoDE}) will be 
simply
    \begin{equation}
\label{rhoDEs22}
\rho_{DE}= \tilde{c}_1 H^2 + \tilde{c}_2 H^{2\gamma+4},
\end{equation}
with $\tilde{c}_1= c_1 \pi M_{p}^{2}$ and  $\tilde{c}_2=  c_2 \left(\pi 
M_{p}^{2}\right)^{-\gamma}$.
We stress here that this scenario does not possess standard holographic dark 
energy as a particular limit in the case where the power-law entropy becomes 
standard Bekenstein-Hawking one (i.e. for $c_2=0$) however it can be a very 
efficient cosmological scenario, being able to describe acceleration. The 
reason is that when the second term in the rhs of (\ref{rhoDEs22}) is present  
the dark energy density obtains richer behavior, which is impossible in the 
case of its absence. Actually it was due to this feature that in the initial 
formulation of holographic dark energy  the event horizon was 
used instead of the Hubble one \cite{Li:2004rb}, however in the present 
power-law extension this complication is not needed, since the Hubble horizon 
can perfectly be used as the IR cutoff.

Interestingly enough, the energy density (\ref{rhoDEs22}) belongs to the class 
of running vacuum models \cite{Shapiro:2009dh,Perico:2013mna}, 
which may reveal the connection between holography and the   renormalization 
group running  \cite{Shapiro:2000dz,Basilakos:2019acj}.
Inserting (\ref{rhoDEs22}) into the Friedmann equation (\ref{Fr1b}) and using 
(\ref{Hrel2}) we obtain 
\begin{equation} \label{OmegaHubble2}
\Omega_{D E}=\tilde{C}_{1}+\tilde{C}_{2}\left[\frac{H_{0}^{2} \Omega_{m 
0}}{a^{3}\left(1-\Omega_{D E}\right)}\right]^{\gamma+1},
\end{equation}
with $\tilde{C}_{1}\equiv \tilde{c_1}/(3M_p^2)$ and $\tilde{C}_{2}\equiv 
\tilde{c_2}/(3M_p^2)$.
Equation (\ref{OmegaHubble2}) is an algebraic equation for $\Omega_{D E}(a)$ 
which can then lead to $\Omega_{D E}(x)$ using  $ 
x=\ln a$. Additionally, the dark-energy equation-of-state parameter can 
straightforwardly arise by inserting (\ref{rhoDEs22}) into the conservation 
equation $\rho_{DE}'+3\rho_{DE}(1+w_{DE})=0$, and eliminating $H$ through 
(\ref{Hrel2}), resulting to  
 \begin{equation} \label{wDE22}
w_{D E}=-\frac{\Omega_{DE}^\prime}{3\Omega_{DE}(1-\Omega_{DE})}.
\end{equation}
Finally, the corresponding deceleration parameter   results inserting the 
above expression into (\ref{qdeccel}).

\section{Cosmological evolution}
\label{Cosmolevolut}

In the previous section we constructed power-law holographic dark energy in a 
consistent way, providing the differential equation that determines the 
evolution of the dark-energy density parameter and then the expression for the 
dark-energy equation of state. In this section we perform a detailed analysis 
of 
the cosmological evolution. We mention that we will focus mainly on Model 1, 
namely of the usual case of the event horizon, since the phenomenology of 
Model 
2 belongs to that of running vacuum scenarios.

\begin{figure}[ht]
\centering
\includegraphics[width=6.cm]{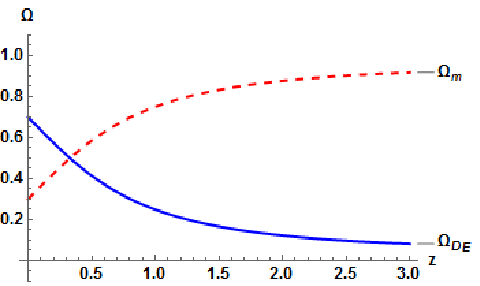}    \\                           
       \includegraphics[width=6.cm]{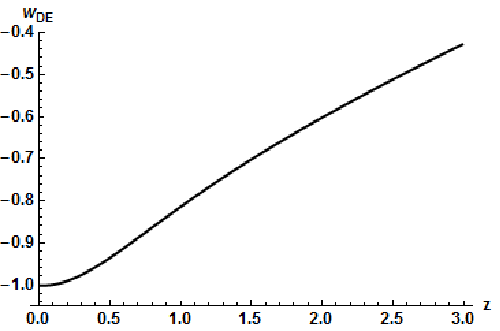} \\
\includegraphics[width=6.cm]{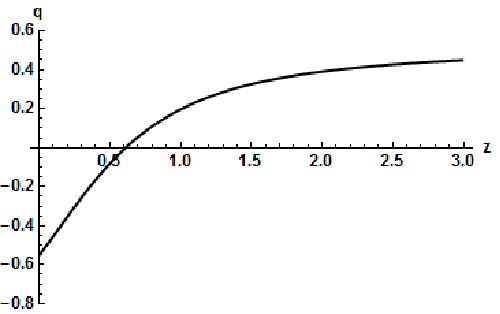}
\caption{\it{ Upper graph: The  power-law holographic  dark energy   
 density   parameter $\Omega_{DE}$ (blue-solid)  and   the matter density
parameter $\Omega_{m}$ (red-dashed), as a function of the redshift $z$, for   
Model 1,   for  $\gamma=0.4$ and
$C_1=0.4$   in units where  $k_{_B}=c=\hbar=M_{p}=1$. 
  Middle graph: The    corresponding dark-energy equation-of-state 
parameter $w_{DE}$. Lower graph:  The   
corresponding   deceleration parameter $q$. In all graphs we have imposed 
 $\Omega_{DE}(x=-\ln(1+z)=0)\equiv\Omega_{DE0}\approx0.7$.
 }}
\label{basicplot}
\end{figure}
 
 In the case of Model 1, namely in the case where the IR cutoff is the   future 
event horizon, since (\ref{difequation})  can be solved 
analytically   only when  $C_2=0$, for the general case we solve it 
numerically. Additionally, in order to be closer to observations, we will use 
the redshift $z$ as the independent variable, which is related to the 
aforementioned variable $x$ through the simple relation  $x\equiv\ln 
a=-\ln(1+z)$. Finally, as usual we impose 
$\Omega_{DE}(x=-\ln(1+z)=0)\equiv\Omega_{DE0}\approx0.7$   
\cite{Planck:2018vyg}.

In Fig. \ref{basicplot} we present the evolution of the  dark energy   and 
matter  density   parameters  (upper graph), of the  dark-energy 
equation-of-state 
parameter $w_{DE}$  (middle graph), and of the  deceleration parameter $q$ 
(lower graph). As we see, we can obtain the  sequence of matter and dark energy 
eras as required by observations, and moreover the  $w_{DE}$ at present is 
around $-1$, while being always in the quintessence regime. Furthermore, $q$  
  exhibits the transition from deceleration to acceleration at  $z\approx 
0.55$, 
  in agreement with the value obtained by $\Lambda$CDM 
scenario \cite{Planck:2018vyg}.

\begin{figure}[ht]
\centering
\includegraphics[width=8.5cm]{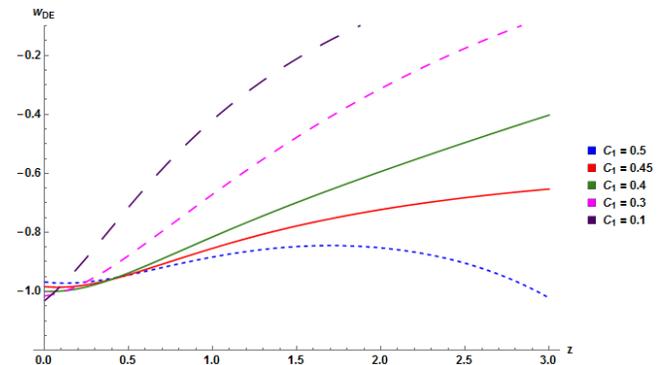}                             
  \caption{\it{ The redshift-evolution of the equation-of-state parameter 
$w_{DE}$ of power-law
holographic dark energy of   
Model 1,      for fixed $\gamma=0.5$ and various values of $C_1$  in units 
where 
 $k_{_B}=c=\hbar=M_{p}=1$. 
We have imposed 
$\Omega_{DE0}\approx0.7$.
}}
\label{wDEparamC}
\end{figure}

\begin{figure}[ht]
\centering
\includegraphics[width=8.5cm]{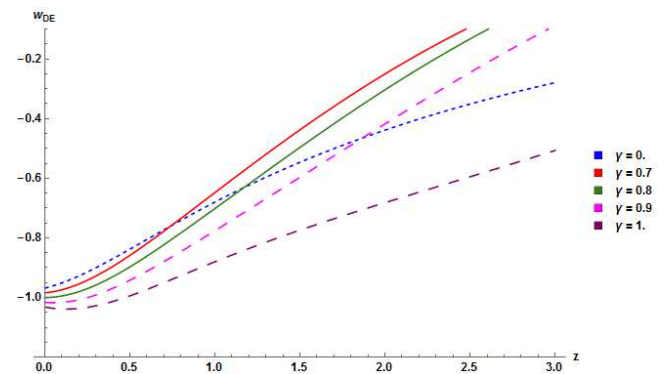}                             
   \caption{\it{ The redshift-evolution of the equation-of-state parameter 
$w_{DE}$ of power-law
holographic dark energy of   
Model 1,      for fixed $C_1=0.35$ and various values of $\gamma$  in units 
where  $k_{_B}=c=\hbar=M_{p}=1$.
We have imposed 
$\Omega_{DE0}\approx0.7$.
}}
\label{wDEparamgamma}
\end{figure}

In order to investigate the effects of the   model parameters $C_1$ and 
$\gamma$ on the cosmological evolution, we focus on the corresponding behavior 
of the dark-energy  equation-of-state  parameter $w_{DE}$.
  In Fig. \ref{wDEparamC}  
we show $w_{DE}(z)$ for fixed $\gamma$ and various values of $C_1$, where for 
all curves the corresponding evolution of the density parameters is similar to 
that of Fig. \ref{basicplot}.
As we observe, with increasing $C_1$ the values of $w_{DE}(z)$ at larger $z$ 
decrease, obtaining phantom values for $C_1>0.5$. On the other hand, the 
present-day value $w_{DE}(z=0)$ remains very close to -1 as required, however 
note that it can lie either in the quintessence regime or in the phantom 
one. Hence, according to the $C_1$  value, the dark-energy  
equation-of-state 
 parameter  can be quintessence-like, phantom-like, or experience the 
phantom-divide crossing during the evolution.
Similarly, in   Fig. \ref{wDEparamgamma} we depict the evolution of 
$w_{DE}(z)$ for fixed  $C_1$ and various values of $\gamma$.
In this case we 
see that for increasing $\gamma$, $w_{DE}$ acquires algebraically larger values 
at larger redshifts, while the present value  $w_{DE}(z=0)$ is around -1, and 
can be either quintessence or phantom.

\section{Conclusions}
\label{Conclusions}
 
In the present work we formulated a  holographic dark energy by considering the 
power-law extended entropy. This entropy relation arises from the     
consideration of  state mixing between the ground and the 
  excited  ones  in the calculation of the entanglement entropy 
\cite{Das:2007mj}. The resulting power-law  holographic dark energy can be 
applied in two horizon choices. In the first model we made the standard choice 
of the future event horizon as the  IR  cutoff, acquiring 
 a one-parameter extension of standard  holographic dark energy, recovering it 
in the limit where power-law extended entropy recovers Bekenstein-Hawking one.
In the second model we used the Hubble horizon as the Universe horizon, and 
although  this scenario does not possess standard holographic dark 
energy as a particular limit, it is a reasonable choice since in this case one 
does not face the usual problems which arise when trying to formulate standard  
holographic dark energy with the Hubble horizon.

We proceeded by extracting the differential equation that determines the 
evolution of the dark-energy density parameter of power-law holographic dark 
energy, and we provided the expression for the corresponding  equation-of-state 
parameter, for both models. Thus, elaborating the equations we are able to 
investigate the resulting cosmological behavior.

Power-law holographic dark energy can describe the sequence of epochs in the 
Universe evolution, namely the domination of matter followed by the domination 
of dark energy, with the onset of acceleration taking place at $z\approx0.55$.
Furthermore, the dark-energy equation of state presents a rich behavior, lying 
in the quintessence regime or passing into the phantom one too, depending on 
the values of the two model parameters. As we saw, for fixed exponent $\gamma$
 with increasing $C_1$ the values of $w_{DE}$ at larger $z$ 
decrease, obtaining phantom values for $C_1>0.5$. On the other hand, for fixed   
$C_1$  increasing $\gamma$ led to algebraically larger values 
at larger redshifts. This behavior is richer than the one of standard 
holographic dark energy, which was expected since the present model has one 
additional parameter.

Before closing this manuscript, let us make a comment on the confrontation with 
observational data. As it is known, the basic holographic dark energy model
is less efficient in fitting the data comparing to $\Lambda$CDM cosmology
\cite{Zhang:2005hs,Huang:2004wt,Wang:2005ph,Feng:2007wn}. 
On the other hand it is free from the 
fine-tuning problem of the latter, and moreover it has been shown to be able to 
 alleviate the $H_{0}$ tension 
 \cite{Abdalla:2022yfr}.  Thus, the field is open, and the 
community still devotes 
research on various extensions of  holographic dark energy   in order to 
examine whether they could be a 
viable candidate for the description of nature.

 Hence, it would be interesting to proceed with a detailed confrontation with 
observations, in order to examine the cosmological capability of 
the scenario of power-law holographic dark energy, and more importantly to 
apply the known information criteria in order to compare its statistical 
efficiency with other scenarios, such as standard holographic dark energy and 
$\Lambda$CDM cosmology. In the same lines, we should study in detail the 
potential ability of the scenario to alleviate the $H_0$ and growth tensions. 
Finally, a necessary step is to perform 
 a detailed investigation of the 
phase-space behavior, through a dynamical-system analysis that could reveal the 
global features of the universe evolution in such a scenario.  These 
investigations lie beyond the present work and 
will be performed in following studies.

    \begin{acknowledgments} 
The authors would like to acknowledge the contribution of the COST Action 
CA18108 ``Quantum Gravity Phenomenology in the multi-messenger approach''.

\end{acknowledgments}

\end{document}